\documentclass[]{elsart}
\usepackage{natbib}
\usepackage{epsfig}
\usepackage{amsmath, amsthm}

\def\apgt{\ {\raise-.5ex\hbox{$\buildrel>\over\sim$}}\ }
\def\aplt{\ {\raise-.5ex\hbox{$\buildrel<\over\sim$}}\ }
\def\lt{\ {\raise-.5ex\hbox{$\buildrel>$}}\ }
\def\gt{\ {\raise-.5ex\hbox{$\buildrel<$}}\ }

\begin{document}

\begin{frontmatter}

\title{High Performance Direct Gravitational N-body Simulations on
Graphics Processing Units\\
\vspace{15pt}
II: An implementation in CUDA}

\author[SCS]{Robert G. Belleman}
\author[SCS]{Jeroen B\'edorf}
\author[SCS,API]{Simon F. Portegies Zwart}

\address[SCS]{Section Computational Science, University of Amsterdam, Amsterdam, The Netherlands}
\address[API]{Astronomical Institute "Anton Pannekoek" , University of Amsterdam, Amsterdam, The Netherlands}

\begin{abstract}

We present the results of gravitational direct $N$-body simulations
using the Graphics Processing Unit (GPU) on a commercial NVIDIA
GeForce 8800GTX designed for gaming computers.  The force evaluation of
the $N$-body problem is implemented in ``Compute Unified Device
Architecture'' (CUDA) using the GPU to speed-up the calculations.  We
tested the implementation on three different $N$-body codes: two
direct $N$-body integration codes, using the 4th order
predictor-corrector Hermite integrator with block time-steps, and one
Barnes-Hut treecode, which uses a 2nd order leapfrog integration
scheme.  The integration of the equations of motions for all codes is
performed on the host CPU.

We find that for $N > 512$ particles the GPU outperforms the
GRAPE-6Af, if some softening in the force calculation is
accepted. Without softening and for very small integration time steps
the GRAPE still outperforms the GPU.  We conclude that modern GPUs
offer an attractive alternative to GRAPE-6Af special purpose hardware.
Using the same time-step criterion, the total energy of the $N$-body
system was conserved better than to one in $10^6$ on the GPU, only
about an order of magnitude worse than obtained with GRAPE-6Af. For $N
\apgt 10^5$ the 8800GTX outperforms the host CPU by a factor of about 100
and runs at about the same speed as the GRAPE-6Af.

\end{abstract}

\begin{keyword}
  gravitation --
  stellar dynamics --
  methods: N-body simulation --
  methods: numerical
\end{keyword}
\end{frontmatter}

\section{Introduction}
The introduction of multiple processing cores in one chip allows
microprocessor manufacturers to improve the performance of CPUs while
the clock rate stays the same. This multi-core principle is not new.
Over the last decade, a similar approach has been taken by
manufacturers of graphics processing units (GPU) under the influence
of the gaming industry to deliver increasingly detailed and responsive
computer games. As a result of this, the GPU underwent a dramatic
increase in performance; a doubling in performance over a period of 9
months, instead of 18 months for CPUs \citep{CUDADOCref,Moore}.

In terms of raw performance, today's GPUs outperform conventional
CPUs. For example, the NVIDIA GeForce 8800GTX
has a performance of about 350\,GFLOP/s (see \S\,\ref{Sect:Results}).
However, harvesting this computing power is not trivial as GPUs are
designed and optimized for graphics operations.  Over the last 7 years
GPUs have evolved from fixed function hardware for the support of
primitive graphical operations to programmable processors that
outperform conventional CPUs, in particular for vectorizable parallel
operations.  Today's GPUs contain many multiple smaller processors
called stream processors \citep{owensStream}, that are specialized in
processing large amounts of data in a streaming and parallel fashion.
It is because of these developments that more and more people use the
GPU for wider purposes than just for graphics
\citep{GPUGems1,GPUGems2,BrookForGPUref}.

Initially, the programming of GPUs was done in assembly language and
required a very specific knowledge of the hardware. Newer generations
of GPUs offered more possibilities for the programmer and with this
came the need for high-level programming languages. With the
introduction of shading languages like Cg \citep{Cgref} and
GLSL \citep{GLSLref}, the programmer could focus on the problem at hand.

Around this time, the performance of the GPU attracted the attention
of researchers with an interest in the use of the GPU as a
high-performance coprocessor. First implementations mapped their
problems into a graphics problem where data is represented as coloured
pixels stored in textures. Shading programs were then used to perform
computations on the data. Although not every problem is easily
represented as a graphics problem, the use of the GPU was demonstrated
in many scientific areas, including but not limiting to PDE solvers,
ray tracing, image segmentation and gravitational simulations
\citep{Owens:2007:ASO}.


One downside of the GPU is that the current generation only supports
single precision (32-bit) floating point arithmetic. This limits their
use to applications for which single precision is sufficient.  In the
release notes of Compute Unified Device Architecture (CUDA) version
0.8, NVIDIA announced that GPUs supporting 64-bit double precision
floating point arithmetic will become available in late 2007
\citep{CUDADOCref}.

In this second paper on high performance $N$-body simulations using
GPUs, we present an implantation using CUDA, and apply the
implementation to solve gravitational $N$-body systems using direct
integration as well as using a Barnes-Hut tree code
\citep{1986Natur.324..446B}.  In our previous paper
\citep{2007astro.ph..2058P} (which we from now on will call ``paper I'') we
presented an implementation in Cg, and showed that for $N \apgt 10^4$
the GPU outperforms the CPU by about a order of magnitude.

The implementation described in this paper was born while we were
drinking beer (whereas \cite{2007astro.ph..3100H} drank tea), so we
have named our implementation {\tt kirin} after a Japanese brand of
beer.  In \S\,\ref{Sect:Nbody} we cover the background of the $N$-body
problem and previous implementations.  Section
\S\,\ref{Sect:Implementation} describes our implementation.  The last
two sections, \S\,\ref{Sect:Results} and \S\,\ref{Sect:Discussion}
cover the results and the discussion.

\section{Background}\label{Sect:Nbody}

The $N$-body gravitational algorithm is based on the force equation as
discovered by Newton. The equation calculates the force between two
particles in space:

\begin{equation}
  {\bf F}_i \equiv m_i {\bf a}_i =     m_i G \sum^{N}
                   _{j=1, j \ne i}
                   m_j
                  {{\bf r}_i-{\bf r}_j \over |{\bf r}_i-{\bf r}_j|^3}.
\label{Eq:Force}\end{equation}

Here $G$ is the Newton constant, $m_i$ is the mass of star $i$ and
${\bf r}_i$ is the position of star $i$. The total force ${\bf F}_i$
(or the acceleration ${\bf a}_i$) that is exercised on particle $i$ is
the summation over the forces between $i$ and all $N$ particles.

In order to determine the total force on each particle within an
$N$-body system, the force exerted by all $N$ particles has to be
calculated. Calculating the force of all particles in the $N$-body
system requires ${1 \over 2} N(N-1)$ force calculations. This
{\large{O}($N^2$)} part of the algorithm is the computationally most
expensive part.  The calculation of the force exerted on each particle
is independent of the calculations performed for other particles. This
makes the calculation of the forces for all particles parallelizable.

A breakthrough in direct-summation $N$-body simulations came in the
late 1990s with the development of the GRAPE series of special-purpose
computers \citep{1998sssp.book.....M}, which achieves spectacular
speedups by implementing the entire force calculation in hardware and
placing many force pipelines on a single chip.  The latest special
purpose computer for gravitational $N$-body simulations, GRAPE6,
performs at a peak speed of about 64\,TFLOP/s
\citep{2001ASPC..228...87M}.  The GRAPE opened the way for the
simulation of large star clusters.  In simulation software such as
{\tt starlab} \citep{2001MNRAS.321..199P}, for example, the GRAPE is
used as a coprocessor for the force calculations.  In this paper we
compare our results with the GRAPE-6Af, which is a smaller commercial
version of the GRAPE6. The GRAPE-6Af contains four GRAPE6 chips that
are mounted on a PCI-card. The performance of the GRAPE-6Af is
123~GFLOP/s and the memory has a maximum capacity of 131072 particles.

Graphics Processing Units (GPU) can be used as an alternative
coprocessor to the GRAPE in $N$-body calculations.  GPUs contain many
processing units that each perform the same series of operations on
different streams of input data, a technique which is better known as
Single Instruction Multiple Data (SIMD).  The first gravitational
$N$-body simulations on GPUs were presented by 
\cite{NylandRef} and later their implementation was improved by Mark
Harris \citep{HarrisRef}. Their implementation only performs force
calculations using a simplified shared time-step algorithm.  A Cg
implementation that performs force, jerk and potential calculations on
a GPU through a block time-step algorithm is described in paper I.
There we concluded that for large $N$ the GPU offers an attractive
alternative for the GRAPE-6Af because of its wide availability, low
price and high reliability.

Recently the use of GPUs has attracted a lot of attention for performing
direct $N$-body force calculations
\citep{2007astro.ph..3100H,ElsenATI2007}.
\cite{ElsenATI2007} uses AMD/ATI hardware, whereas
\cite{2007astro.ph..3100H} uses NVIDIA GPU cards.
The latter also use CUDA to implement the force calculations,
achieving an even higher performance than presented in paper I. 
\cite{2007astro.ph..3100H} tested the code only on an
implementation using shared time-steps and with softening.
We present a library, implemented in CUDA, that uses similar principles
as the implementation by \citep{2007astro.ph..3100H}. Our library
(called {\tt kirin}) can
be used for direct $N$-body simulations as well as for treecodes,
it can be run with shared-time steps or with block time-steps and
allows non-softened potentials.

The CUDA framework exposes the GPU as a parallel data streaming
processor that consists of many processing units. Compared with
previous programming interfaces such as Cg, CUDA provides more
flexibility to efficiently map a computing problem onto the hardware
architecture.  CUDA applications consist of two parts. The first
executes on the GPU and is called a ``kernel''. Kernels are
implemented in the CUDA programming language, which is basically the
``C'' programming language extended with a number of keywords.  The
other part executes on the host CPU and provides control over data
transfers between CPU and GPU and the execution of kernels.

A kernel program is run by multiple threads that run on the GPU.  We
call a group of threads a bundle.  Threads contained in the same
bundle communicate with each other using shared memory and cannot
communicate with threads in another bundle.  Calculations on the GPU
are started by specifying the number of bundles to execute and the
number of threads that each bundle contains. The total number of
threads is the product of the two.

The NVIDIA GeForce 8800GTX hardware architecture defines a
hierarchical memory structure where each level has a different size,
access restrictions and access speed.  In general, accessing the
largest type of memory is flexible but slow, while accessing the
smallest type of memory is restrictive but fast.  This memory
structure is directly exposed by the CUDA programming framework.  The
challenge in mapping a computing problem efficiently on a GPU through
CUDA is to store frequently used data items in the fastest memory,
while keeping as much of the data on the device as possible.

Current GPUs support 32-bit IEEE floating point numbers, which is
below the average general purpose processor, but for many applications
it turns out that the higher (double) precision can be emulated at
some cost or is not crucial.
The relatively low accuracy of the GPU hinders high precision direct
$N$-body integrations, but is very suitable for methods which
intrinsically have a lower precision, such as the Barnes-Hut
treecode. We therefore tested our library for GPU-enabled $N$-body
simulations on a direct integration method (\S\,\ref{Sect:testcode}
and \S\,\ref{Sect:ResWithKira}) as well as using a treecode
(\S\,\ref{Sect:treecode}).

\section{Implementation}\label{Sect:Implementation}

The $N$-body scheme used in our implementation is described by 
\cite{1992PASJ...44..141M}. The integration
scheme consists of three parts: a predictor step that predicts a
particle's position and velocity; a Hermite integrator to advance the
position and velocity to the new time-step and a corrector step that
corrects the predicted position and velocity using the results of the
integrator.  The acceleration, its time derivative (jerk) and
potential are computed by direct summation.

\subsection{Decomposition over CPU and GPU}\label{Sect:decomposition}
In our implementation, the calculation of force, potential and jerk is
performed on the GPU. The predictor and corrector steps are performed
on the CPU.  Our algorithm uses a block time-step scheme that only
integrates subsets (blocks) of particles that need to be updated
\citep{1993ApJ...414..200M}.

The decomposition of this scheme over a CPU and GPU was done for two
reasons. First; the prediction and correction steps are more sensitive
to round-off errors and are therefore best performed using the CPU's
64-bit floating point representation. Second; production quality
software such as {\tt starlab} \citep{2001MNRAS.321..199P} uses a
similar decomposition, but then in combination with the GRAPE
coprocessor.  We opted for a similar decomposition as used for the
GRAPE to allow astronomical production software to link in our GPU
implementation as a library.

Our implementation requires that particle data is communicated between
the CPU and the GPU at each block time-step.  This is accomplished
through a number of memory copies where the CPU sends particle
position, velocity and mass to the GPU.  The results computed by the
GPU (acceleration, jerk and potential) are retrieved by the CPU.
For the GPU library the prediction is performed on the CPU after which
all particles are copied to the GPU.  The GRAPE only has to send the
updated particles and performs the prediction on the GRAPE hardware
itself. This results in an overall lower performance for the GPU than
for the GRAPE, because the overhead of the memory copies increases
much more for the GPU than for the GRAPE.

The input and output variables exchanged with the GPU program are the
following:
\begin{itemize}
  \item{} Input:  mass ($N$), position ($3N$) and velocity ($3N$),
  \item{} Output: acceleration ($3N$), jerk ($3N$) and potential ($N$)
\end{itemize}
All values are represented by single precision (32-bit) floating point
values, which is the most precise representation offered by current
generation GPUs. This adds up to 14~floats or 56~bytes per particle
which results in a total capacity of approximately 14~million
particles for the 768MB on-board memory available on the GeForce
8800GTX.  This is a substantial increase in capacity compared to the
GRAPE-6Af's maximum capacity of 131072 particles.  This is also an
improvement over the 9~million particles that could be stored with the
earlier Cg implementation in paper I.  A restriction
imposed by Cg that does not allow memory areas to be readable and
writable at the same time forced this implementation to use a
double-buffering scheme.  This restriction does not exist in the CUDA
implementation described here.

The fundamental structure of our CUDA implementation aims at
exploiting the available computing resources as much as possible.  The
challenge in mapping our $N$-body problem on a GPU through CUDA is to
annihilate wait states due to slow memory accesses while keeping the
threads executing on the GPU occupied.

Global memory access is slow (400 to 600 clock cycles)
 while shared memory access is fast (4 clock cycles) but
has a limited capacity. We therefore pre-cache particles into shared
memory up to its maximum capacity before the calculation of forces.
The input data is split in smaller parts that are each
pre-cached and processed in consecutive bursts.

The integration of one block time-step is initialised by assigning a
thread to each of the particles in a block.  Each thread then goes
through the following steps:

\begin{enumerate}
\item{} Each thread in the bundle caches one particle from global
memory into the shared memory.  The total number of read particles is
therefor equal to the number of threads contained in a bundle.

\item{} The force, potential and jerk for the thread are calculated
using the particles that are cached in shared memory.  The thread then
sums the partial results into temporary variables which are stored in
a register.

\item{} Steps (1) and (2) are repeated until all $N$ particles have been
read.

\item{} When all parts are processed, the self interaction of the
potential value is removed, the results are saved in global memory and
the thread exits.

\end{enumerate}

Note that the total number of calculations performed by the GPU with
this scheme is $N^2$.  Although it is possible to determine the force
using ${1 \over 2} N(N-1)$ calculations, this would require internal
communication and synchronization. This added communication is costly
in a GPU and would result in lower performance even though less work
is done.

The method of giving each thread its own specific data and allowing
data that is needed by multiple threads to be stored in shared memory
is generally accepted as the best method to reduce memory latency when
using CUDA capable GPUs.  Shared memory significantly reduces the wait
time that occurs while using global memory.  This speeds-up the
algorithm by reducing the number of global memory accesses.

\begin{figure}
  \center \includegraphics[width=0.5\textwidth]{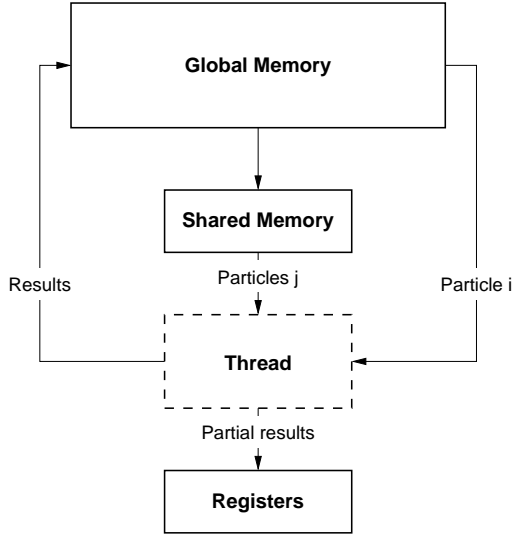}
  \caption{The memory access strategy used in our implementation to
  compute the force for particle $i$. Data for particle $i$ is taken
  directly from global memory. Each kernel copies data for particle
  $j$ from (slow) global memory into (fast) shared memory in parts
  and performs the calculations on particle $j$. This is repeated for
  all particles. Intermediate results are stored in registers.}

  \label{fig:Memory}
\end{figure}

In our implementation the number of bundles that is started depends
 on the number of
particles in the current time-step block. Each bundle in our
implementation contains 128 threads. Therefore the force, jerk and
potential of 128 particles is calculated in parallel. In comparison;
the GRAPE-6Af does the same but for 48 particles. The number of
bundles that are started is equal to the number of particles in the
time-step block divided by 128. This reduces the number of global
memory accesses by a factor 128.  Our implementation uses the thread
scheduler to swap in threads that have already loaded their data while
threads that are waiting on memory loads are swapped out.  Once all
threads have loaded the particle data from global memory into the
shared memory space of the bundle, all threads in the same bundle can
operate on that data.  Through this strategy, the latency incurred by
global memory accesses is hidden, which speeds up the algorithm
considerably. In Fig. \ref{fig:Memory} we illustrate the memory configuration used
in our implementation.




\subsection{Optimizing GPU utilization}
The implementation described in \S\,\ref{Sect:decomposition} has the
disadvantage that it does not utilize all processors in the GPU when
the number of particles in a block time-step is smaller than 4096.
This number is derived as follows: To make full use of all 16
multiprocessors in the GPU it is necessary to start at least 16
bundles. Moreover, threads in a bundle that are waiting for data from
global memory will be swapped out in favour of bundles for which the
data is ready and can be processed, which brings the total number of
bundles to 32.  With our implementation, where we start 128 threads
for each bundle, we must have at least $32 \times 128 = 4096$
particles in the block time-step to fully utilize all 16
multiprocessors.

To fully utilize the GPU for any number of particles in the block
time-step we have altered the implementation in such a way that it
splits the calculations in several parts and then
combines the partial results on the host. This is done when there are
less then 4096 particles in the block time-step.

The implementation divides the total number of particles in several
parts that are processed sequentially. Each part contains 128
particles, equal to the number of threads per bundle. One by one the
threads in each bundle load a particle $i$ from global memory and then
process the particles $j$ that have been loaded in shared memory.  When we
have less than 4096 particles in the block time-step, the parts that
have to be processed are evenly distributed, as much as possible, over
multiple bundles.  Each bundle calculates a partial force between its
particle $i$ and the particles $j$ in the part(s) that have been loaded
from global memory. The partial results are then saved in global
memory. This strategy assures that all multiprocessors in the GPU are
fully utilized. As threads in different blocks cannot communicate it
is not possible to aggregate partial results from finished blocks.
Therefore the partial results are saved in global memory and
subsequently combined on the host CPU. The host CPU loads the partial
results from the GPU and then adds the partial results together.

\subsection{Mimicking the GRAPE6 library}
We have designed a library around our GPU based $N$-body
code that mimicks the standard GRAPE6 library. This allows
existing applications that are linked to the GRAPE6 library to be used
with {\tt kirin} with minimal changes.  Additional requirements are
that the CUDA run-time libraries are installed on the system and that
a graphics card capable of running CUDA applications is installed in
the system. Appendix \ref{sect:g6functions} shows a list of functions that
have a GPU equivalent. GRAPE6 functions that do not require a GPU
equivalent are implemented as dummy functions.

\subsubsection{Kernel changes}

In addition to force, jerk and potential the GRAPE hardware also
calculates the nearest neighbour of every particle that is being
updated, and the GRAPE has the ability to perform calculations without
softening. The softening parameter $\epsilon$, introduced by
\cite{1963MNRAS.126..223A}, prevents very small integration steps
when particles reside very close to each other.  The GPU code has to
be adjusted to calculate the nearest neighbour and to handle
simulations without softening.

Nearest neighbours are determined by comparing the distance between
each particle and all other particles in the data set. This is done as
part of the force calculation; a comparison is added with each force
calculation to maintain the particle with the minimum distance.  When
the force calculation is complete, the index to the nearest neighbour
is saved in global memory, together with the force, jerk and potential
results.

The distance $r_{ij}$ between two particles $i$ and $j$ can be zero
either when $i = j$ or when the distance between two particles
cannot be represented within the limited precision of a single
precision floating point number.  This results in a
division by zero in the force equation.  The softening is added to
the distance and has the effect that the distance between two
particles can never be zero.  For zero softening
the resulting division by zero is circumvented by an additional check
in the inner loop of the GPU program.

Adding each of these two comparisons results in lower performance:
one extra comparison results in a performance
drop of roughly 10\%.  This is mainly caused by the underlying SIMD
architecture that enforces that when two threads take different
branches, one has to wait until the branching thread reaches the same
point in the program as the other. In Appendix \ref{sect:g6functions}
we present a list of the implemented {\tt kirin} library functions.

\section{Results}\label{Sect:Results}

The simulations for the direct integration are run over 0.5 
$N$-body time units \citep{HM1986}\footnote{See also {\tt
http://en.wikipedia.org/wiki/Natural\_units\#N-body\_units}.}, but the
measurements are from $t = 0.25$ to $t = 0.5$ to minimize the effect
of initialization on the measurements. The simulations for the treecode
are run over 1 $N$-body time unit, with the time measurements for 
$t = 0$ to $t = 1$.
The host hardware we used are Hewlett-Packard xw8200
workstations with two Intel Xeon CPUs running at 3.4 GHz. These
machines either had an NVIDIA GeForce 8800GTX graphics card in the PCI
Express ($16\times$) bus or a GRAPE-6Af.  The GRAPE and Cg machines
ran a Linux SMP kernel version 2.6.16, Cg version 1.4 and graphics
card driver 1.0-9746. The {\tt kirin} measurements were performed with
Linux SMP kernel version 2.6.18, CUDA Toolkit version 0.8 and graphics
card driver 1.0-9751.

We compare the energy of the simulated system at the start and end of
the simulation.  The total energy $E$ within an isolated system must
remain constant.  We determine the relative error ${\triangle E/E}$
using the following equation:

\begin{equation}
  {\triangle E/E} =  {E_{start}-E_{end} \over {E_{start}}}. \label{Eq:Error}
\end{equation}

\subsection{Direct $N$-body integration in a test environment }\label{Sect:testcode}

\begin{figure}
  \includegraphics[width=\columnwidth]{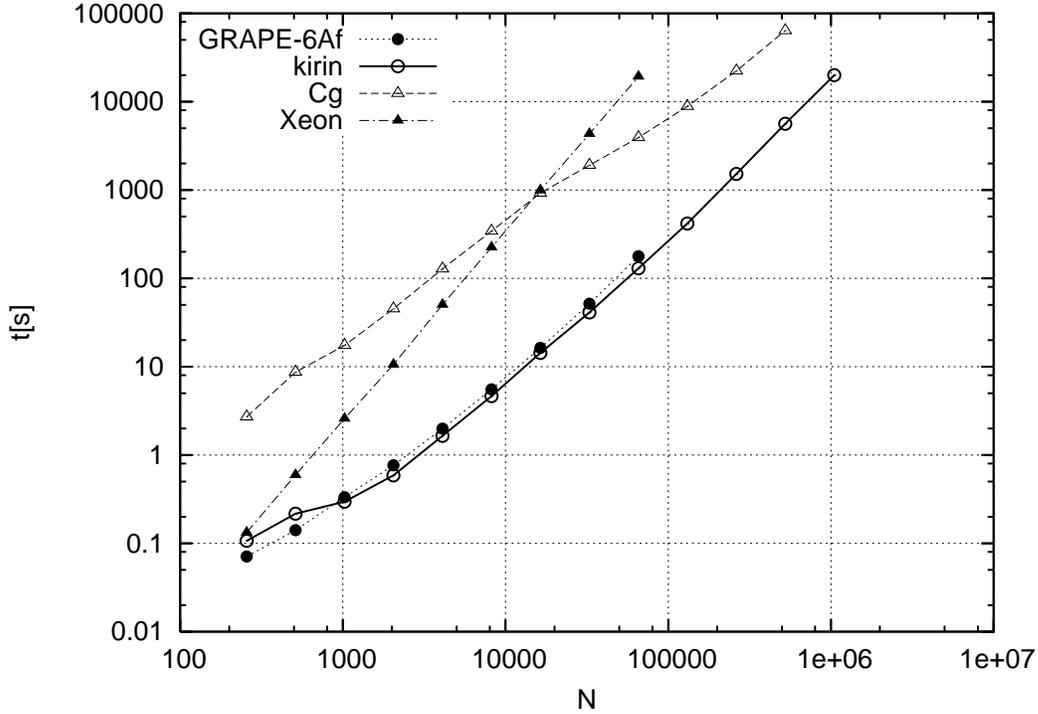}
  \caption{Performance comparison of the $N$-body implementations from
  Table~\ref{Tab:Results}.  {\tt kirin} is
  represented by the solid line (open circles). The GRAPE is
  represented as the dotted line (bullets). The Cg implementation is
  represented as the dashed line (open triangles). The dashed-dotted
  line (closed triangles) represent the results on the host computer.}

  \label{fig:GPU}
\end{figure}

In Table \ref{Tab:Results} we compare the performance of our CUDA
implementation with the GRAPE-6Af hardware and the Cg implementation
described in paper I. Softening is set to
$\epsilon = {1/256}$ to enable comparison with other implementations
(\cite{2006NewA...12..169N} and paper I). Later in
\S\,\ref{Sect:ResWithKira} we relax this assumption.  In Fig.
\ref{fig:GPU} we have plotted the performance of the different
implementations.  In Table~\ref{Tab:Error} we present the measurements
of the error ${\triangle E/E}$.

\begin{table}
\caption[]{Performance of {\tt kirin} compared to other
implementations.  The first column ($N$) gives the number of equal
mass particles of a Plummer sphere. Columns 2 to 5 show the
performance of the different implementations.  The GRAPE-6Af column
shows the result on GRAPE hardware.  {\tt kirin} and the Cg
implementation ran on the NVIDIA GeForce 8800GTX.  The last column
shows the performance of an implementation that ran completely on the
host, an Intel Xeon at 3.4 GHz. The simulations were run over
$0.5$ $N$-body time unit (timing measurements were done from $t =
0.25$ to $t=0.5$). The softening parameter used is 1/256.
Some measurements are performed for limited $N$
for practical reasons. The results on the GRAPE are limited to up to
65536 because of a defective memory chip.  }

\label{Tab:Results}
\begin{tabular}{rccccccc}
\hline

$N$       & GRAPE-6Af   & {\tt kirin} & Cg   & Xeon  \\
         &       $[s]$      & $[s]$ & $[s]$ & $[s]$  \\
\hline
256       &     0.07098 &    0.130       &  2.708   &  0.1325\\
512       &     0.1410&    0.359        &  8.777   &  0.5941 \\
1024      &     0.3327&    0.297        &  17.46   &  2.584 \\
2048      &     0.7652&    0.588        &  45.27   &  10.59 \\
4096      &     1.991&    1.646        &   128.3   &  50.40 \\
8192      &     5.552&    4.631        &   342.7   &  224.7 \\
16384     &     16.32&    14.28        &   924.4   &  994.0 \\
32768     &     51.68&    41.16        &   1907    &  4328  \\
65536     &     178.2&    129.8        &   3973    &  19290 \\
131072    &     -   &    417.6            &   8844    &  -  \\
262144    &     -   &    1522            &   22330   &   -  \\
524288    &     -   &    5627             &   63960   &   - \\
1048576   &     -   &    19975            &     -     &   - \\
\hline
\end{tabular}
\end{table}

\begin{table}
\caption[]{The relative energy error $\triangle E/E$ of the various
implementations. The first column ($N$) gives the number of equal mass
particles of a Plummer sphere.  Columns 2 to 4 show $\triangle E/E$ of
the different implementations. The relative error was obtained by
running the simulation over 0.5 $N$-body time unit with $\epsilon =
1/256$ using the same input parameters as used in the measurements for
Table~\ref{Tab:Results}. }

\label{Tab:Error}
\begin{tabular}{rrrrcc}
\hline
$N$       & GRAPE  & {\tt kirin} &  Cg     \\
        &   $[\times 10^{-7}]$      &  $[\times 10^{-7}]$ &  $[\times 10^{-7}]$ \\
\hline
256       &    2.271&   3.554        &  3.554   \\
512       &    2.388&   1.209        &  2.419  \\
1024      &    0.866 &   2.375      &  -8.909  \\
2048     &    1.261  &   2.366       &  -35.500     \\
4096     &    -1.881 &   -1.204	   &  -4.815         \\
8192      &    2.560 &   3.609	        &  25.261         \\
16384     &    -0.818&   -1.189	   &  61.840          \\
32768     &    -1.363 &   -1.898	  &  24.986         \\
65536     &    -6.150 &   -4.767      &  2.383          \\
131072    &      -&   22.634           &  195.790        \\
262144    &      -&   26.147          &  -118.850 \\
524288    &      -&   80.482           &  -164.450 \\
1048576   &      -    &  -116.552      &     -    \\
\hline
\end{tabular}
\end{table}

We also measured the peak performance of our implementation by
disregarding the communication between host and GPU; only the actual
calculations are timed.  The results shown in
Table~\ref{Tab:FResFlops} give the performance measurements when
calculating only the force. The results in Table~\ref{Tab:FJResFlops}
give the performance measurements when calculating force, potential
and jerk.  The performance ($P$) in floating point operations per
second (FLOP/s) is calculated using:

\begin{equation}
  {P} = kN^2 / t.
\label{Eq:Flops}\end{equation}

\noindent
Here $k$ is the number of floating point operations used in the
calculations. We use $k=38$ for the force calculation. This
number was introduced by Warren et al.
\citep{10.1109SC.1997.10057} and is used as reference number in other
papers \citep{2006NewA...12..169N,2007astro.ph..3100H}. For
calculating force, potential and jerk we use $k=60$, as used by
Makino et al. in \cite{2006NewA...12..169N,2006astro.ph..6105N}.

The numbers in Table \ref{Tab:FResFlops} indicate a peak performance of
340~GFLOP/s. The theoretical peak performance of the 8800GTX is
346~GFLOP/s.\footnote{The 8800GTX has 128 processing units at 1350
MHz. Each can execute 2 instructions at the same time (multiply and
add). This results in $1350 \times 128 \times 2 = 345.6$ GFLOP/s.}
This means we have practically reached the theoretical peak speed of
the GPU.

\begin{table}
\caption[]{Peak performance measurements when calculating only the
force. The first column indicates the number of particles.  The second
column shows the execution time for {\tt kirin}. The third column
shows the performance in GFLOP/s calculated using equation
\ref{Eq:Flops} with $k=38$. The fourth and fifth columns give the same
results for the Chamomile scheme described in
\cite{2007astro.ph..3100H}.}

\label{Tab:FResFlops}
\begin{tabular}{rcrclr}
\hline
$N$       &  {\tt kirin} &  Speed & \vline & Chamomile   & Speed \\
          &  $[s]$ &  GFLOP/s    &   \vline & $[s]$ & GFLOP/s \\
\hline
256       &   0.000090  &    27.46     & \vline &  -  &     - \\
512       &   0.000091  &    109.0    & \vline & -  &     - \\
1024      &   0.000180  &    221.2    & \vline & -  &       -  \\
2048      &   0.000537	&    296.6    & \vline &   0.000921  &     173  \\
4096      &   0.001976	&    322.6    & \vline &   0.00299  &     213  \\
8192      &   0.007739	&    329.5    & \vline &   0.01082  &     235  \\
16384     &   0.030205	&    337.7    & \vline &   0.0414  &     246  \\
32768     &   0.122863	&    332.1    & \vline &   0.162  &     251  \\
65536     &   0.479895  &    340.1    & \vline &   1.642  &     254  \\
131072    &   1.9182  &    340.3    & \vline &   2.548  &     256 \\
\hline
\end{tabular}
\end{table}

\begin{table}
\caption[]{Peak performance measurements when calculating force,
potential and jerk. The first column indicates the number of
particles.  The second and third column show the execution time and
performance in GFLOP/s calculated using equation \ref{Eq:Flops} with
$k=60$.}

\label{Tab:FJResFlops}
\begin{tabular}{rrrccr}
\hline
$N$       & {\tt kirin}   & Speed\\
          &  $[s]$ &  GFLOP/s    \\
\hline
256       &    0.000132  &     29.78 \\
512       &    0.000133  &     117.93 \\
1024      &    0.000336  &     187.24  \\
2048      &    0.001149  &     219.02  \\
4096      &    0.004416  &     227.95  \\
8192      &    0.017537  &     229.59  \\
16384     &    0.070002  &     230.07  \\
32768     &    0.279824  &     230.23  \\
65536     &    1.118900  &     230.31 \\
131072    &    4.468939  &     230.65 \\
262144    &    17.87493  &     230.67 \\
524288    &    71.51776  &     230.61 \\
1048576   &    279.4067  &     236.11 \\
\hline
\end{tabular}
\end{table}

\subsection{Direct $N$-body integration in a production environment}\label{Sect:ResWithKira}


We have linked our library with the integrator that is used in the
{\tt starlab} software package ({\tt kira}). The {\tt kira} integrator
has built-in support for the GRAPE6 hardware and therefore no code
changes besides renaming the {\tt G6\_} functions were needed. 

\begin{figure}
  \includegraphics[width=\columnwidth]{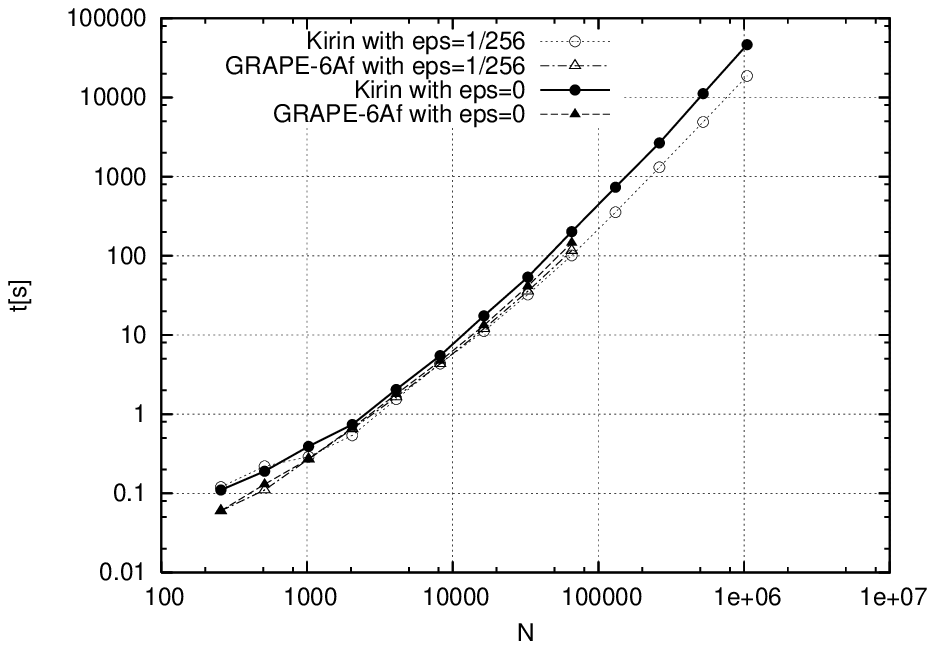}
  \caption{Performance comparison of the $N$-body implementations from
  Table~\ref{Tab:ResultsKira}, using the {\tt kira} integrator in {\tt
  starlab}. The {\tt kirin} library  with $\epsilon = 1/256$ is
  represented by the dotted line (open circles).  The same library
  with $\epsilon = 0$ is represented by the solid line (bullets).  The
  standard GRAPE6 library with $\epsilon = 1/256$ is represented by
  the dash-dotted line (open triangles).  The same library with
  $\epsilon = 0$ is represented by the dashed line (closed
  triangles).}

  \label{fig:ResultsKira}
\end{figure}

The {\tt starlab} simulation results are found in Table
\ref{Tab:ResultsKira}.  We compare the performance of the GPU with the
GRAPE6-Af. We have performed simulations for a range of data sets
starting with $N=256$ up to $N= 1048576$ (The GRAPE results are
limited to $N=65536$).  The simulations are run over 0.25 $N$-body
time-unit.  We have used two different softening values, namely
$\epsilon=1/256$ as we have used in the test environment Section
(\S\,\ref{Sect:testcode}) and $\epsilon=0$.  The used accuracy
parameter is 0.3 (The ``a'' parameter in {\tt starlab} which
controls the time step).
In Fig.  \ref{fig:ResultsKira} we have
plotted the performance of the GPU and of the GRAPE.  The relative
errors of the simulations can be found in Table \ref{Tab:ErrorKira}.

\begin{table}
\caption[]{Performance measurements comparing execution time of the
standard GRAPE6 library with our GPU library. The test are performed
by using the {\tt starlab} software package.  Columns 2 and 3 show the
GRAPE and GPU results with $\epsilon= 1/256$. Columns 4 and 5 show the
results of the same simulation, but now with $\epsilon=0$.}
\label{Tab:ResultsKira}
\begin{tabular}{rccccc}
\hline
          & $\epsilon= 1/256$ &         & \vline & $\epsilon=0$ & \\
          \hline
$N$       & GRAPE-6Af   & {\tt kirin}   & \vline & GRAPE-6Af   & {\tt kirin} \\
        &   $[s]$      & $[s]$          & \vline &   $[s]$      & $[s]$ \\
\hline
256       &     0.06    &    0.12    & \vline & 0.06 & 0.11\\
512       &     0.11    &    0.22    & \vline & 0.13 & 0.19\\
1024      &     0.27    &    0.29    & \vline & 0.27 & 0.39\\
2048      &     0.65    &    0.54    & \vline & 0.67 & 0.74\\
4096      &     1.65    &    1.51    & \vline & 1.79 & 3.75\\
8192      &     4.33   &    4.35     & \vline & 4.7 & 8.57\\
16384     &     12.02   &    11.17   & \vline & 13.18 & 20.2 \\
32768     &     35.69   &    32.5    & \vline & 41.4 & 57.1 \\
65536     &     116.1     &  101.1   & \vline & 146 & 202 \\
131072    &     -     &    355       & \vline & - & 735\\
262144    &     -   &    1313     & \vline   & - & 2668\\
524288    &     -   &    4913     & \vline  & - & 11190\\
1048576   &     -   &    18681    & \vline  & - & 46372\\
\hline
\end{tabular}
\end{table}

\begin{table}
\caption[]{The relative energy error $\triangle E/E$ of the simulations performed
with {\tt kira}. The first column ($N$) gives the number of equal mass
particles of a Plummer sphere.  Columns 2 and 3 show $\triangle E/E$
for the GRAPE and the GPU using $\epsilon=1/256$. Columns 4 and 5 show $\triangle E/E$
for the GRAPE and the GPU using $\epsilon=0$. The relative error was obtained by
running the simulation over 0.25 $N$-body time unit using the same
input parameters as used in the measurements for
Table~\ref{Tab:ResultsKira}. }
\label{Tab:ErrorKira}
\begin{tabular}{rccccccc}
\hline
          & $\epsilon= 1/256$ &         & \vline & $\epsilon=0$ & \\
          \hline
$N$       & GRAPE-6Af   & {\tt kirin}       & \vline  & GRAPE-6Af   & {\tt kirin} \\
&  $[\times 10^{-7}]$ &  $[\times 10^{-7}]$ & \vline & $[\times 10^{-7}]$ &  $[\times 10^{-7}]$ \\
\hline
256       &     1.14    &    0.4      & \vline &   -0.105 & -2.0\\
512       &     0.331   &    -0.397   & \vline &   0.734  & -0.0128\\
1024      &     -0.253  &   -0.78    & \vline  &   -0.53  & -0.908\\
2048      &     0.213   &    0.31    & \vline  &   0.156  & 0.126\\
4096      &     -8.71   &    -8.92   & \vline  &  -10.09  & -11.6\\
8192      &     -51.5   &    -51.5   & \vline  &   -151   & -151\\
16384     &     -3.75   &    -3.46   & \vline  &  -86.1   & -86.2\\
32768     &     8.32    &    8.14    & \vline  &   497    & 4.98\\
65536     &     37.0    &   37.3   & \vline    &   1420   & 1413\\
131072    &     -       &    28.5   & \vline   &   -      & 188\\
262144    &     -       &    15.9     & \vline &   -      & 2606  \\
524288    &     -       &    -40.4    & \vline &   -      & 7582  \\
1048576   &     -       &    -94.2    & \vline &   -      & 5789 \\
\hline
\end{tabular}
\end{table}

\subsection{$N$-body integration using the treecode}\label{Sect:treecode}

\begin{figure}
  \includegraphics[width=\columnwidth]{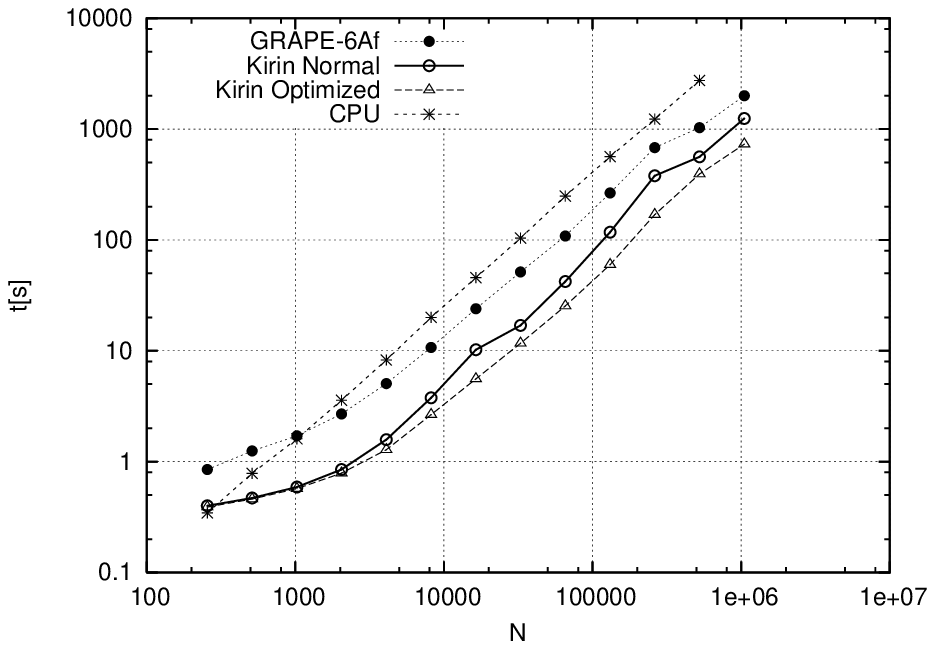}
  \caption{Performance comparison of the execution time of the
  treecode from Table~\ref{Tab:ResultsTree} over 1 $N$-body time unit.
  The GRAPE hardware is represented with the dotted line (bullets),
  the normal version of the GRAPE mimicking library is represented as
  the solid line (open circles). The optimized version of the library
  is represented as the dashed line (open triangles).  The CPU is
  represented as the dashed line (stars).}

  \label{fig:ResultsTreecode}
\end{figure}

We have applied our {\tt kirin} library to run with the treecode
\citep{1986Natur.324..446B} as implemented by
\cite{2004PASJ...56..521M}. This implementation has been designed to
run on a GRAPE. Therefore we have linked the source code with our
library to let the algorithm run on the GPU.  The results of these
calculations, run on GRAPE, GPU and CPU, are presented in
Table \ref{Tab:ResultsTree}. In Fig.\,\ref{fig:ResultsTreecode} we
have plotted the performance of the different implementations.

We adapted two different implementations of the library, the first is
identical to the one described in \S\,\ref{Sect:ResWithKira}, the
second one is optimized for the treecode.  The Barnes-Hut treecode
algorithm performs time integration using acceleration only, we
therefore can leave out the jerk and nearest neighbours calculations.
This results in a performance gain of a factor of two (see
Fig.\,\ref{fig:ResultsTreecode}).  The direct integration method
requires, besides the acceleration, also the derivative of the
acceleration (jerk). Besides the jerk the {\tt kira} integrator also
requires the nearest neighbour of each particle that is integrated.
Since the jerk and the nearest neighbour are not needed for the
integration using the treecode we can disable the code that calculates
the jerk and the nearest neighbour to get extra performance.  The
relative errors of the simulations can be found in Table
\ref{Tab:ErrorTree}.


\begin{table}
\caption[]{Performance measurements comparing the execution time of
the treecode using the standard GRAPE6 hardware, the GPU and the CPU.
For the GRAPE and GPU we choose an ``ncrit'' value of either 8192,
16384 or 32768; whichever was fastest (the ``ncrit'' value controls
the average number of particles in a group).  Other than this, all
simulations are run over 1 $N$-body time unit with default settings.}

\label{Tab:ResultsTree}
\begin{tabular}{rccccc}
\hline
$N$       & GRAPE-6Af   & {\tt kirin (normal)} &  {\tt kirin (optimized)} & CPU \\
        &   $[s]$      & $[s]$  & $[s]$ & $[s]$\\
\hline
256       &     0.85    &    0.40       &   0.39 &   0.34    \\
512       &     1.25    &    0.47        &  0.46 &   0.78  \\
1024      &     0.71    &    0.59        &  0.57 &   1.61   \\
2048      &     2.69    &    0.85        &  0.79 &   3.58   \\
4096      &     5.07    &    1.58        &  1.28 &   8.27   \\
8192      &     10.7   &    3.77        &   2.65 &   19.9 \\
16384     &     23.9   &    10.2        &   5.57 &   45.6 \\
32768     &     51.4   &    16.9        &   11.7 &   104    \\
65536     &     109     &    42.3        &   25.4 &  249     \\
131072    &     266     &    117         &   59.9 &  564      \\
262144    &     682   &    379           &   169 &   1230  \\
524288    &     1033   &    563           &  394 &   2752   \\
1048576   &     2004   &    1247            &     733 &  5985      \\
\hline
\end{tabular}
\end{table}

\begin{table}
\caption[]{The relative energy error $\triangle E/E$ of the simulations performed
using the treecode algorithm. The first column ($N$) gives the number of equal mass
particles of a Plummer sphere.  Columns 2 to 4 show $\triangle E/E$
for the GRAPE, GPU and CPU respectively. The relative error was obtained by
running the simulation over 1 $N$-body time unit using the same
input parameters as used in the measurements for
Table~\ref{Tab:ResultsTree}. }
\label{Tab:ErrorTree}
\begin{tabular}{rccccc}
\hline
$N$       & GRAPE-6Af   & {\tt kirin} & CPU \\
&  $[\times 10^{-6}]$ &  $[\times 10^{-6}]$ & $[\times 10^{-6}]$\\
\hline
256       &     496    &   496    & 345   \\
512       &     3.41  &      3.46   & 545 \\
1024      &     8.03      &  8.02   & 122 \\
2048      &     5.19    &    5.17   & 876 \\
4096      &     6.78   &     6.78   & 592 \\
8192      &     5.76   &     5.80   & 217  \\
16384     &     0.126   &    0.08   & 300  \\
32768     &     25.4  &     25.4    & 32.0 \\
65536     &     66.7     &  66.8    & 145\\
131072    &     42.2     &    42.3  & 70.0  \\
262144    &     29.9   &    30.2    & 38.8   \\
524288    &     13.2   &    13.2    & 13.1  \\
1048576   &     17.8   &    18.0    & 19.1 \\
\hline
\end{tabular}
\end{table}

\section{Discussion}\label{Sect:Discussion}


The use of graphics processing units offers an attractive alternative
to specialised hardware, like GRAPE.  While GPUs are programmable,
however limited, they can be deployed for a wider range of problems,
whereas GRAPE is single purpose. Also the cost for purchase and
maintenance of a GPU is much lower than for GRAPE.  However, the
single precision of current GPUs remains a problem, as we already
stated in paper I.  Note also that the GRAPE we used is the smallest
1-module PCI version, and obviously we cannot outperform a TFLOP/s
GRAPE-6 board of the full 64 TFLOP/s GRAPE system with a single GPU.

In Fig.\,\ref{fig:GPU} we compare the performance of GRAPE-6Af with
the GPU.  For small system of particles ($N \aplt 512$),
GRAPE remains superior in speed by about a factor of two when
integrating the equations of motion using the block time-step
scheme.


For systems with $N>512$ our implementation in CUDA
performs at comparable speed as the GRAPE-6Af. For such a large number
of particles, most block time-steps utilise the GPU at full capacity.
The earlier implementation in Cg (paper I) is slower
by about a factor ten compared to {\tt kirin}.


The performance of {\tt kirin} depends on the amount of bundles and
threads that are started. Since the optimal number of threads and
bundles depends on the design of the GPU, it is hard to provide an
optimal value. The maximum number of threads that can be initialised
cannot exceed the number of registers available to store the partial
accelerations, jerks and potentials.  The overall performance depends
therefore on the number of registers available on the multiprocessors.
Ideally CUDA should have a routine that returns the optimal number of
threads and bundles.

In our implementation the performance of {\tt kirin} increases from $N
= 256$ to reach almost peak performance at $N \simeq 4096$.  For
larger number of particles, the performance hardly increases, as in
these cases the GPU is fully utilised (see Table
\ref{Tab:FJResFlops}).  In Table\,\ref{Tab:FResFlops} we compare the
performance of {\tt kirin} with the recently published {\tt Chamomile}
scheme \citep{2007astro.ph..3100H}.  It is interesting to note that the
latter scheme shows the same scaling behaviour as our implementation,
though about 35\% slower than {\tt kirin}. The comparison in
Table\,\ref{Tab:FResFlops}, however, shows a situation in
which only the forces are calculated, without calculation of the
higher derivatives that are needed for the
Hermite integration scheme. Ignoring the jerk and potential
calculations allows more threads to be initialised as fewer registers
will be occupied.

In Table \ref{Tab:FJResFlops} we present the performance measurements
for calculating the force, the potential and the jerk on the GPU. This
performance is lower than those presented in
Table\,\ref{Tab:FResFlops}, but the jerk and potential is
needed for a more accurate integration of the equations of motion.
The maximum performance we obtain using a GPU is about 230\,GFLOP/s.

In Fig.\,\ref{fig:ResultsKira} we compare the performance of the
GRAPE-6Af with the GPU.  For $N > 512$ and  $\epsilon$ = ${1/256}$,
our {\tt kirin} library performs with a comparable performance as the
GRAPE-6Af. Without softening the integration steps are smaller which
results in a lower performance of our {\tt kirin} library than the
GRAPE-6Af.  The relative error in the energy of the GRAPE and the GPU
are of the same order for both softening values as can be seen in
Table \ref{Tab:ErrorKira}.

Reducing the accuracy of the integrator in the calculations with GRAPE
results in a linear response to the computation time. Increasing the
accuracy with a factor of two results in an increase in the
computation time of a factor of two, but a decrease in the energy
error of a factor of $2^4$. Increasing the accuracy while running on
the GPU with a factor of two results in an increase in the computation
time of about an order of magnitude, whereas the energy error hardly
decreases.

In Fig.\,\ref{fig:ResultsTreecode} we compare the performance of our
library implementation with the GRAPE and the CPU for the treecode.
The performance scaling is roughly the same for the GPU, CPU and the
GRAPE, except that the GPU implementation is an order of magnitude
faster than the CPU implementation.  The treecode sends all particles
to the hardware during each time-step. The number of memory copies to
the GRAPE is the same with the GPU. As a consequence the GPU
outperforms the GRAPE for all $N$ because we are not limited by the
memory transfers.  The relative error in the energy of the treecode is
comparable for the GRAPE and the GPU for all $N$.

Throughout our simulations, both the GPU and the GRAPE produce a
relative error in the energy of the order of $|\Delta E|/E \sim
10^{-7}$, over a range of $N=256$ to 65536 particles, which is
consistent with the results in paper I. Reducing
the integration time steps will result in a smaller error for the
GRAPE while the GPU error stays more or less the same
\citep{2007astro.ph..2058P}. We expect that the introduction of double
precision GPUs later in 2007 will result in a better conservation of
the energy, and if this will not affect performance too negatively,
GPUs will become a real challenge to GRAPE.


At the moment it is impractical to implement the predictor and
corrector part of the integration scheme on the GPU, mainly because of
the limited precision. The future double precision hardware may
resolve this problem, in which case we can expect an even greater
speedup for the GPU supported $N$-body simulations, in particular
since it would reduce the communication between the GPU and the host
computer. An example of this can already be partially seen in the
treecode results where we outperform the GRAPE because less memory
transfers are required.

\section*{Acknowledgements}

We are grateful to Mark Harris and David Luebke of NVIDIA for
supplying us with the two NVIDIA GeForce 8800GTX graphics cards on
which part of the simulations were performed. This work was supported
by NWO (via grant \#635.000.303 and \#643.200.503) and the Netherlands
Advanced School for Astrophysics (NOVA).  The calculations for this
work were done on the Hewlett-Packard xw8200 workstation cluster and
the MoDeStA computer in Amsterdam, both are hosted by SARA Computing
and Networking Services, Amsterdam.

\bibliographystyle{aa}
\bibliography{biblibrary}

\newpage
\begin{center}
{\bf APPENDIX}
\end{center}

\appendix
\section{{\tt kirin} library functions}\label{sect:g6functions}

The {\tt kirin} library is compatible with the GRAPE6 library. As a
result, all existing code that uses the GRAPE6 library only needs to
be recompiled and relinked to use the GPU equivalents of the GRAPE6
functions. All functions in the GRAPE6 library have an equivalent
GPU implementation. The most important are listed below:

\begin{itemize}
\item{}
{\tt GPU\_open} - opens the connection with the GPU and initializes local
buffers.
\item{}
{\tt GPU\_close} - closes the connection with the GPU and releases all allocated memory (local as well as
on the GPU).
\item{}
{\tt GPU\_npipes} - returns the number of pipelines that are on the
chip (for the GRAPE this is 48).  The GPU does not have a fixed number
of pipelines, therefore the number can be configured using a
configuration file. Tests show that for some applications the code is
slowed down if the number of pipes is set too high.


\item{}
{\tt GPU\_set\_j\_particle} - sets a particle in the memory of the GPU. The particle
will first be stored in a local buffer and then sent to the GPU after the prediction step.
\item{}
{\tt GPU\_set\_ti} - sets the next time step to be used by the predictor,
starts the predictor on the host system and sends the predicted particles to the GPU.

\item{}
{\tt GPUcalc\_firsthalf} - has the same effect on the GPU and GRAPE; The force calculation for the particles specified in the function call will be started.
\item{}
{\tt GPUcalc\_lasthalf} - has the same effect on the GPU and GRAPE; The results
of the previous {\tt GPUcalc\_firsthalf} call will be retrieved.
\end{itemize}

\end{document}